\documentclass[%
 reprint,
 amsmath,amssymb,
 aps,
]{revtex4-2}

\usepackage{graphicx}
\usepackage{dcolumn}
\usepackage{bm}

\usepackage{units}

\usepackage{tikz}

\usepackage{lipsum}

\begin{document}

\preprint{APS/123-QED}

\title{Geometrical fingerprints of shear transformation zones in network glasses}

\author{Franz Bamer}
 \email{bamer@iam.rwth-aachen.de} 
\author{Zhao Wu}
\author{Somar Shekh Alshabab}
\affiliation{%
 RWTH Aachen University \\
 Eilfschornsteinstraße 18, 52062 Aachen, Germany
}%


\date{\today}

\begin{abstract}
Oxide glasses have the structure of disordered covalent networks that are accountable for their mechanical response.
Identifying the topological phenomena of the elastic structural response, we statistically backpropagate
local regions that have the highest susceptibility of rearrangement. Shear transformation zones in network glasses highly correlate with regions of the highest variance in their bond stretch distributions projected into the direction of macroscopic deformation. However, directional influence is significantly less essential than bond stretch variance, which shows that shear transformation zones in network glasses are mainly state-dependent. Exclusively depending on the local geometry of the initial material state, our indicators are physically informing and can be evaluated directly from images with insignificant computational effort.
\end{abstract}

\maketitle



Inelastic deformation, such as plasticity or fracture, occurs during structural changes in the atomic configuration. Although the deformation of inelastic material response can be highly complex, it originates from elementary mechanisms whose nature depends on the material type.
For ordered structures, that is, crystals, the mechanism of inelastic deformation is relatively well understood, where plasticity evolves around clearly detectable defects embedded in the otherwise well-ordered atomic structure \cite{Li2010,Takeuchi2021}. Thus, the material response depends on the location of these defects and their particular arrangement with each other.
Although the material response of disordered structures, such as glasses, varies strongly from ordered structures, it also originates from point defects that experience local atomic scale rearrangements \cite{Falk1998,Tsamados2009}. Driven by mechanical deformation, the material rearranges locally in these point defects, while the surrounding matrix responds elastically so that defects may interact during catastrophic, avalanche-type events leading to complex mechanisms such as plastic deformation or fracture \cite{Maloney2004a,Maloney2004b,Maloney2006}.

Unfortunately, for disordered solids, the appearance of such defects, is not apparent from observing their local structural picture before mechanical deformation. Thus, significant effort has been spent on finding such local spots and predicting their mechanical activation \cite{Richard2020,Bamer2023}.
In particular, detecting zones prone to rearrangement includes the incorporation of soft vibrational modes
\cite{Tanguy2010,Ding2014,Tyukodi2016,Yang2020}, nonlinear plastic modes \cite{Gartner2016,Kriuchevskyi2022}, local mechanical probing \cite{Patinet2016,Barbot2018,Ruan2022}, finding adjacent minima investigating the potential energy landscape \cite{Xu2018}, investigating nonaffine deformation fields \cite{Xu2021}, and machine learning-based strategies \cite{Cubuk2015,Fan2021,Font-Clos2022}. Recently, Hardin et al.~\cite{Hardin2024} have developed a generalized distance function, the Gaussian Integral Inner Product distance, as a low-dimensional local descriptor of disordered solids. Since the softness of local spots may also depend on the deformation protocol \cite{Gendelman2015}, Schwartzman-Nowik et al.~\cite{Schwartzman-Nowik2019} proposed a predictor depending on the local heat capacity and the linear response of the material subjected to external deformation.
Introducing purely geometrical quantities as predictors, that is, indicators that do not include dynamical quantities, such as potential energies and their derivatives, or data-driven black-box analyses, such as neural networks, are particularly attractive since they do not rely on additional assumptions from molecular simulations and provide meaningful physical interpretations. A popular and simple purely geometric predictor in this regard is the local free volume \cite{Spaepen1977,Spaepen2006}, which is based on the idea that free volume must be available to be occupied by particles during a possible local rearrangement. Based on these ideas, Rieser et al.~\cite{Rieser2016} proposed a local characterization by local anisotropy using Voronoi cells.
In this letter, we propose a class of purely structural predictors that are particularly suitable for disordered network materials. We present approaches that depend on the local anisotropy and its alignment with the macroscopic deformation protocol as well as local bond stretching.


        

\begin{figure}
    \centering
    \includegraphics[width=1.0\linewidth]{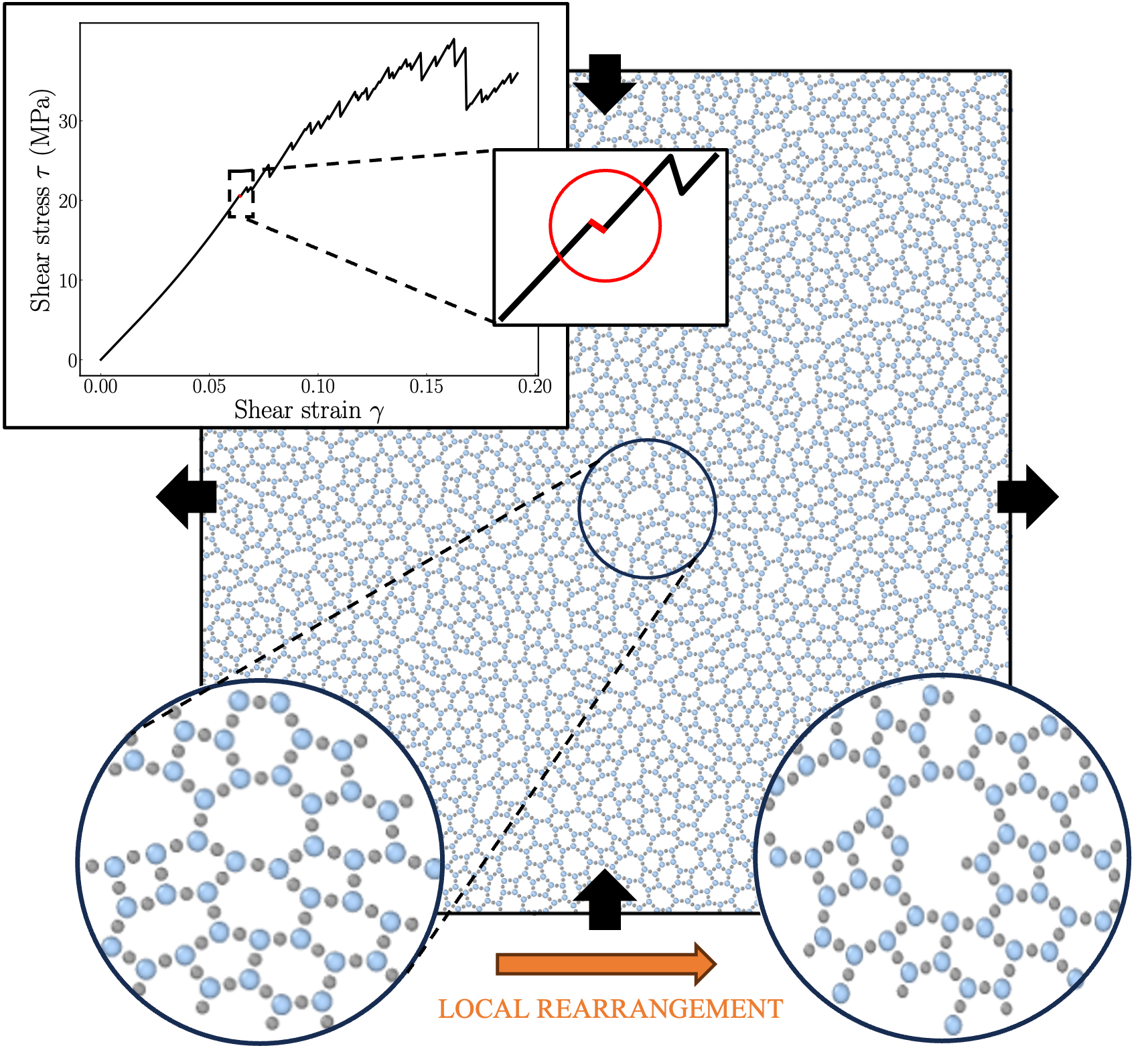}
    \caption{Network glass sample subjected to a pure shear loading protocol. Local rearrangements in shear deformations occur as bond braking events where the topology of the material alters.}
    \label{fig:network_glass_sample}
\end{figure}

\textit{Mechanical model ---}Our mechanical benchmark systems are numerical realizations of two-dimensional silica glass images \cite{Lichtenstein2012,Huang2012,Huang2013}. These flat material patches are bilayer structures that are mirrored in the out-of-plane direction. The in-plane, two-dimensional image information provides the essential properties for our numerical network glass models, which follow the topological constraints of a Zachariasen glass \cite{Zachariasen1932}. They are particularly useful for our investigations since they allow for a direct visual inspection while having network structures that are experimentally verified via atomic imaging.
Starting from a hexagonal lattice configuration with periodic boundary conditions, we perform a Monte Carlo bond-switching algorithm, elaborately discussed in Bamer et al.~\cite{Bamer2023}. This strategy is a Monte Carlo Markov chain approach where the switching sequence is performed based on an objective function that quantifies the topological difference of the switched sample with target network statistics from the imaged benchmark sample \cite{Ormrod_morley2018}. Hereby, the topology is quantified by the overall statistics of ring sizes and the ring neighborhood statistics using the Aboav-Weaire law \cite{Bamer2023}.
Our network glass samples are the result of $2e4$ consecutive switching attempts one of which is shown in Figure \ref{fig:network_glass_sample}a. The interaction of this two-dimensional SiO$_3$ model glass is modelled by a Yukawa-type potential using a cutoff radius of $10$ \r A, elaborated on in Roy et al.~\cite{Roy2018}.
We generated a set of $50$ network glass samples of $1e4$ atoms whose topological structure is statistically equivalent to the benchmark sample imaged by Lichtenstein et al.~\cite{Lichtenstein2012}.
Mechanical deformation was performed by elongating the rectangular simulation cell in the direction of the first Bravais vector while compression the simulation cell in the direction of the second Bravais vector and ensuring that the area of the cell remains unaltered. This way, one performs pure shear loading with the principal components of shear aligning with the horizontal and vertical direction. The athermal quasistatic deformation protocol was performed by applying incremental steps of deformation followed by minimizing the potential energy \cite{Maloney2006}. Thus, the structure remains in a minimum of the potential energy landscape, and thermal vibrations are omitted, allowing one to focus purely on the structural response. The corresponding shear stress-strain curve of the sample is also shown in the top inlay of Figure \ref{fig:network_glass_sample}. The material responds in elastic branches intersected by sudden stress drops, revealing a typical response of a disordered system. The size of the incremental shear step was chosen as $5.0e\! -\! 4$, which leads to a sufficiently high resolution to detect all stress drop events. Bonfanti et al.~\cite{Bonfanti2019} have detected two types of elementary events in silica glass, that is, angle-changing events that do not alter the topological structure and bond-breaking events that alter the topological structure. In our two-dimensional study, we exclusively observe events of the second type in which one or more covalent bonds in the network rupture, leading to the formation of nanovoids and dangling bonds. An enlarged local cutout of the configuration shortly before and shortly after one such stress drop event is presented in the bottom of Figure \ref{fig:network_glass_sample}.
Our predictions build on the concept that covalent networks mainly distribute external loading through the covalent graph structure in the network. Following this line of thought, (i) bond directions are relevant, and, in particular, their alignment with the external deformation protocol, as well as (ii) the length of the bonds indicating how much an atomic pair in one respective bond can be strained during macroscopic deformation.

\begin{figure}[h!]
    \centering
    \begin{tikzpicture}
        \node at (0,0) {
            \includegraphics[width=0.95\linewidth]{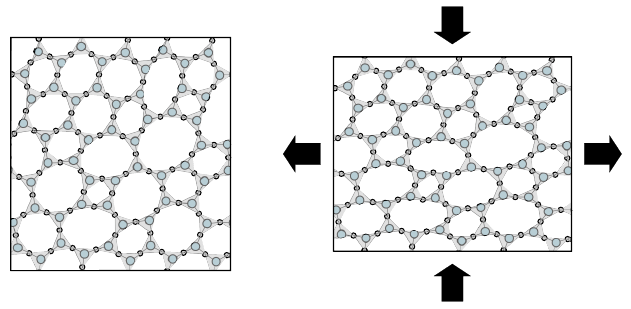}
        };
        \node at (0,-6){
            \includegraphics[trim=24 0 4 0, clip, scale=0.95]{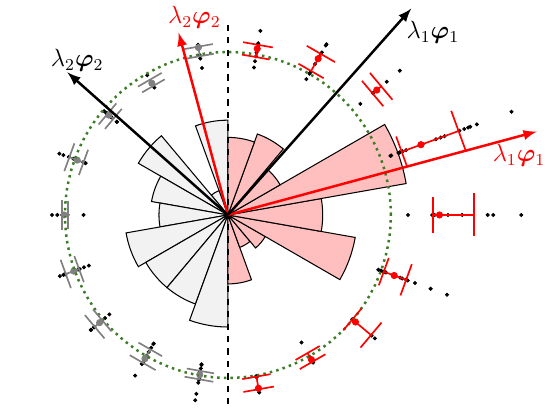}
        };
        \node at (0.0,-12.0){
            \includegraphics[trim=5 0 5 0, clip]{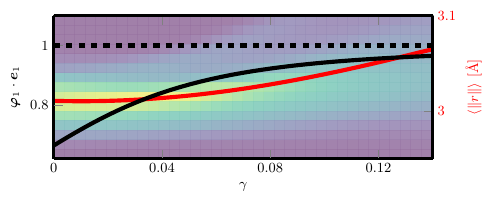}
        };
        \node at (-0.4,2.1){\textbf{(a)}};
        \node at (-0.4,-2.3){\textbf{(b)}};
        \node at (-0.4,-10){\textbf{(c)}};
    \end{tikzpicture}
    \caption{(a) Undeformed network glass sample and elastically deformed network glass sample shortly before an inelastic event; (b) orientation distribution function of the bonds of the undeformed and the elastically deformed sample; (c) projection of the first principal direction of the second-order fabric tensor into the first principal direction of true shear. The heatmap in the background shows the probability density function of the bond lengths at each time step during the simulation, where the yellow color indicates the higher values.}
    \label{fig:network_directions}
\end{figure}
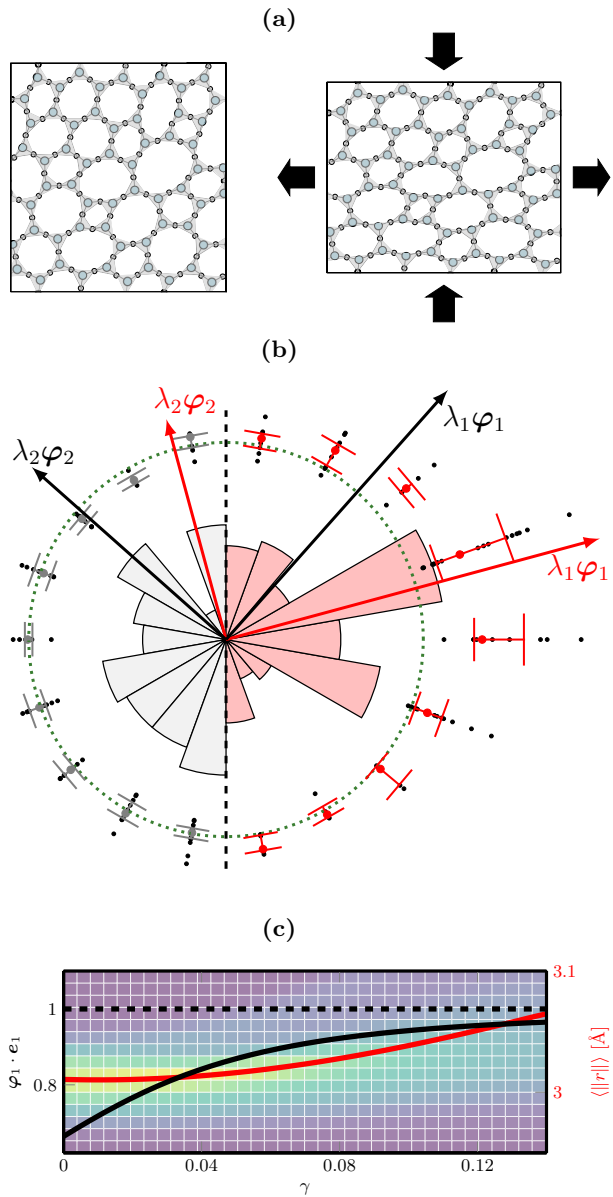

\textit{Deformability of local regions ---}To break down the general mechanism of how network glasses respond to external deformation, we present an illustrative example of $200$ atoms in Figure \ref{fig:network_directions}. This Zachariasen model glass consists of corner-sharing SiO$_3$ triangles forming the graph, which manifests itself as an arrangement of rings of various shapes and sizes. Thus, the silica network is equivalently identified by considering the triangle centers as nodes and the corner-sharing oxygen bonds as the edges of the graph. We perform an athermal pure shear deformation protocol by deforming the configuration in small incremental steps followed by minimization of the potential energy.
The atomic configuration in its initial undeformed state is depicted in Figure \ref{fig:network_directions}a while the deformed configuration is shown in Figure \ref{fig:network_directions}b. The corner-sharing SiO$_3$ triangles are presented in light gray. Since the topology does not change from the initial state in Figure \ref{fig:network_directions}a to the deformed state in Figure \ref{fig:network_directions}b, this deformation is purely elastic. Notably, the deformed configuration in Figure \ref{fig:network_directions}b shows a state shortly before the material becomes locally unstable and drops into an adjacent minimum in the potential energy landscape.
From visual inspection, one observes that the bonds in the undeformed structure are rather evenly distributed in the sample while they align with macroscopic deformation after the elastic loading protocol.
To quantify this deformation-induced anisotropy, we define a random variable for the bond directions $\bm{n}$, with $\|\bm{n}\|=1$, and collect the set of its realizations $\bm{n}^{(\alpha)}$. The orientation distribution function is written as: $\phi(\bm{n}) := \frac{1}{N} \sum_{\alpha=1}^{N} \delta(\bm{n}-\bm{n}^{(\alpha)})$. The histograms of the orientation distribution functions of the unstrained and strained samples are presented in Figures \ref{fig:network_directions}b in gray at the left-hand side and in red at the right-hand side, respectively.
The moments of this distribution are defined by:
\begin{equation}
    N_{i_1i_2...i_n} := \frac{1}{N} \sum_{\alpha=1}^{N} n^{(\alpha)}_{i_1} n^{(\alpha)}_{i_2}...n^{(\alpha)}_{i_n} = \langle n_{i_1} n_{i_2}...n_{i_n} \rangle \; .
\end{equation}
Since we do not define any preferred bond direction, the orientation distribution function is point symmetric, and all fabric tensors of odd order vanish. To emphasize our point, we evaluate the second order fabric tensor, $N_{i_1i_2} = \langle n_{i_1} n_{i_2} \rangle$, and compute its eigenvectors to quantify the level of anisotropy of the system. As shown at the left-hand side of Figure \ref{fig:network_directions}b in gray and at the right-hand side of Figure \ref{fig:network_directions}b in red, the eigenvectors of the material bond directions change significantly from the undeformed to the deformed state.

Inelastic events occur as ruptures of one or more bonds in the network structure, as indicated in Figure \ref{fig:network_glass_sample} and further discussed in Bamer et al.~\cite{Bamer2020}. Figure \ref{fig:network_directions}b also shows the collection of the bond lengths depending on their directions for the initial and deformed states, in gray and red dots, respectively. The corresponding quartiles are represented in terms of black and red lines and the data are expanded by a factor of ten around the average bond length. While the bond length variations are independent of the direction in the undeformed state, the variations significantly increase for the bonds that are aligned into the first principal direction of remote deformation. However, we recall that the absolute number of bonds aligning with the pulling direction of macroscopic deformation increases with external loading.

Figure \ref{fig:network_directions}c shows the evolution of the first eigenvector projected into the first principal direction of true shear in black. With increasing elastic strain, the material gradually aligns with external deformation, and the structure becomes increasingly more anisotropic. At the same time, the red line of this figure shows the bond length statistics with increasing external loading, revealing that the bonds slowly align with the pulling direction and gradually increase in length with external deformation. Furthermore, the average bond length becomes less pronounced leading to a wider distribution with increasing shear deformation.

The network response is initially dominated by bond alignments with the external deformation and gradually transitions into the stretching of bonds that align most with external deformation. Following these findings, we divide the network structures into zones with a larger capacity to align and stretch, which we refer to as flexible regions and zones with a lower capacity to align and stretch, which we refer to as inflexible regions.
In other words, the bonds in the flexible regions are less aligned with the first principal direction of true shear, so they realign during loading without subjecting the individual bonds to significant axial tensile loading.
Covalent bonds belonging to inflexible regions respond instantaneously to mechanical loading since they have no capacity to align and stretch. Such regions transfer the mechanical load through the material sample, while flexible regions deform without load transfer. Consequently, inflexible regions have less capacity to further absorb mechanical loading so that they are more prone to coincide with soft spots or shear transformation zones, which experience bond rupturing events shown in Figure \ref{fig:network_glass_sample}.

\textit{Local geometric predictors ---}To quantify if local regions are prone to experiencing atomic rearrangements in the form of bond-breaking events, we examine circular material cutouts while ignoring the remaining material sample. The radius of these material windows is chosen as $9$ \r A for our problem, resulting in local regions of less than a hundred atoms. The choice of this window size depends on the material and requires rigorous test simulations to find this optimum, as presented in the supplementary material. The objective is to find a predictor $\chi$ that indicates the susceptibility of experiencing a mechanically induced instability by one real positive scalar grading value. The smaller the predictor, the higher the susceptibility to a local bond rearrangement.

We scanned $50$ network glass samples of $10^4$ atoms each using a scanning grid size of $50\!\times\! 50$, resulting in a prediction map $\chi\!\left(x_1,x_2\right)$ discretized by $2500$ points. To assess the quality of our predictors, we first define $F\!\left(\chi_{(\cdot)}\right)$ to be the cumulative distribution function of the respective predictor $\chi_{(\cdot)}$. Then, we identify the sequence of stress drop events so that the predictive quality of every event $n$ is assessed by $\mathcal{C}_{\chi_{(\cdot)}}\!(n) = 1-2F\!\left(\chi_{(\cdot)}\right)$. This way, a value of $1$ stands for a perfect prediction, a value of $0$ indicates no predictive benefit, and a value of $-1$ indicates perfect anticorrelation. All predictors are quantified for the first $20$ events. After that, the material is in a fractured state where large voids are present, and percolation effects dominate during rupture.

We developed and characterized 18 purely structural predictors. However, due to similarities and correlations, we will break these 18 predictors into five types and compare them with the predictive power of the local free volume, which may be seen as a purely structural benchmark predictor in the literature \cite{Falk1998,Spaepen2006}. For the sake of completeness, the remaining predictors are presented in the supplementary material.

\begin{figure}
     \centering
    \includegraphics[width=\linewidth]{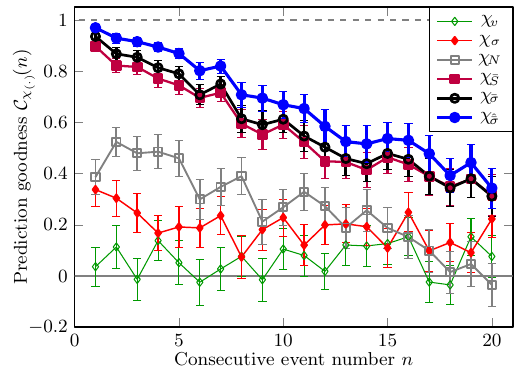}
    \caption{Prediction goodness of purely geometric indicators.}
    \label{fig:prediction_results}
\end{figure}

The local free volume is evaluated by performing a Voronoi tessellation with the development points being the atomic coordinates. In our two-dimensional framework, the local free volume is the sum of the area of the Voronoi cells minus the atomic areas, that is, $v_{\!f} = v_{\text{vor}}-\text{v}_{\text{at}}$. The free volume predictor is defined as being inversely proportional to the free volume, that is, $\chi_{\!v} := \nicefrac{1}{v_{\!f}}$. This way, the assumption is: the larger the free volume, the more susceptible a region is to experiencing atomic scale rearrangements. The evaluation of $\chi_{\!v}$ is shown in green in Figure \ref{fig:prediction_results} for the first $20$ events.
Clearly, the free volume is a bad indicator for locating shear transformation zones in network glasses since no significant difference from the zero axis is achieved, which is not of any predictive merit here since it is equivalent to guessing.
Our first predictor is independent of the deformation protocol and quantifies the local level of disorder. In a two-dimensional framework, network glasses consist of corner-sharing SiO$_3$ triangles, sharing rings of various shapes and sizes. Since the crystalline polymorph of two-dimensional network glasses is an arrangement of six-membered rings in the form of honeycomb lattices, every deviation from this order may be quantified by the standard deviation $\sigma$ of the histograms of the ring statistics taken from every local region. This first predictor is inversely proportional to the level of disorder, that is, $\chi_{\!\sigma}:= \nicefrac{1}{\sigma}$. In other words, the assumption is that the higher the network disorder of a local region is, the more susceptible it is to experiencing atomic-scale rearrangements. The predictor $\chi_{\!\sigma}$ is presented by the red line in Figure \ref{fig:prediction_results}. The variance of the local ring statistics indicates regions prone to rearrangements, albeit, with a prediction performance that is relatively moderate.
In what follows, we focus on the edges of the network graphs of the SiO$_3$ tetrahedra. We start with the initial alignment phase during deformation, shown in Figure \ref{fig:network_directions}b and \ref{fig:network_directions}c and exclusively focus on the bond directions. This way, we collect all bond vectors in the network and normalize their length. The hypothesis is that bonds already aligned with the first principal direction of pure shear are affected earlier by the pure shear deformation protocol and are more likely to break. A similar idea was presented by Reiser et al.~\cite{Rieser2016} who measured the level of anisotropy in metallic glass performing Voronoi tesselation. The predictor is defined by $\chi_{\!N}:= \nicefrac{1}{(\lambda_1\bm{\varphi}_1\cdot\bm{e}_1})$, where $\bm{\varphi}_1$ is the largest eigenvector of the second order fabric tensor $\bm{N}$ and $\bm{e}_1$ denotes the unit vector pointing into the first principal direction of pure shear. The result is shown by the gray plot in Figure \ref{fig:prediction_results}. Although this predictor also indicates regions prone to rearrangements to some extent, its performance is comparable to that of the local network disorder.

\begin{figure}[ht!]
    \centering
    \begin{tikzpicture}
        \node at (0,0){
            \includegraphics[width=\linewidth]{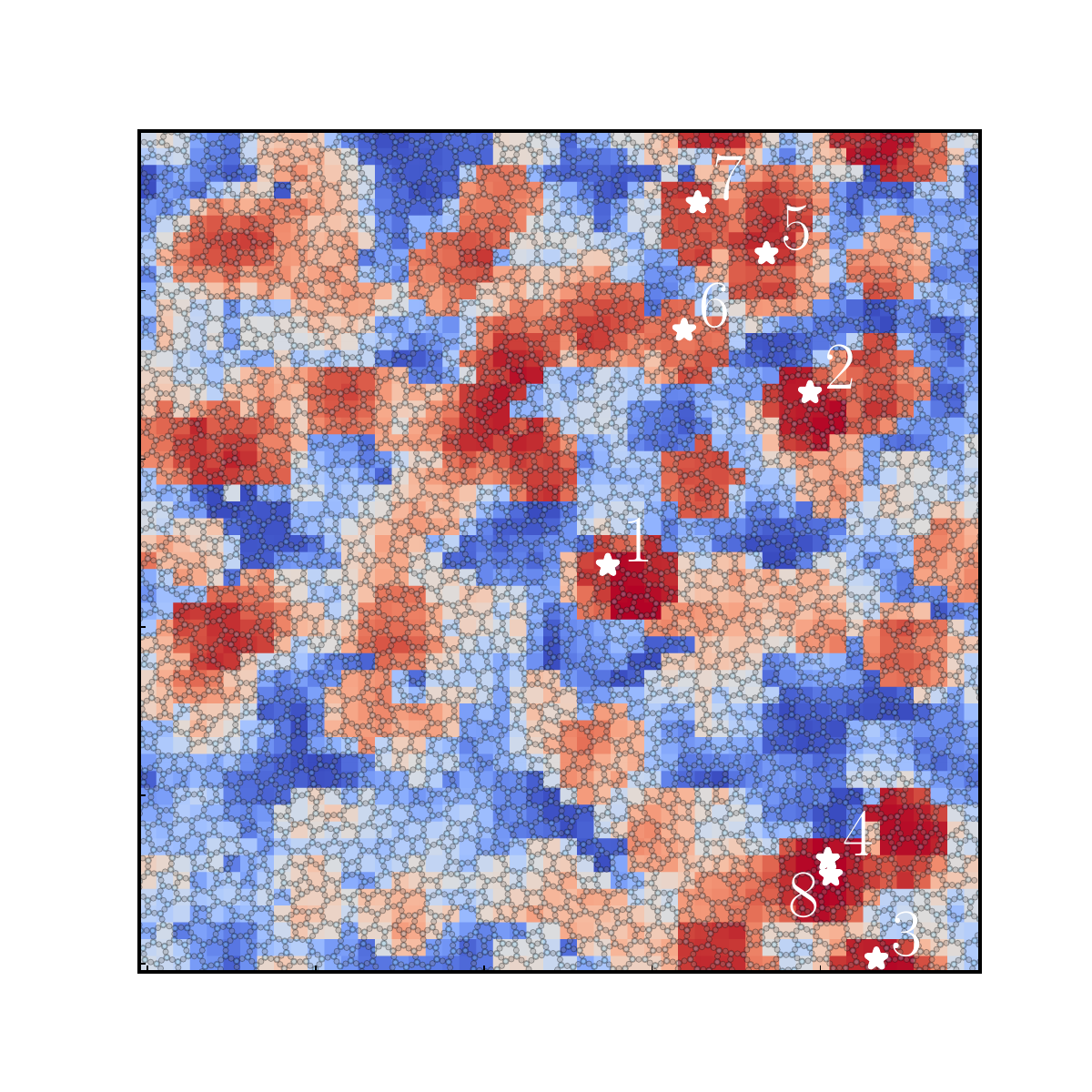}
        };
        \node at (0,-9.45){
            \includegraphics[width=\linewidth]{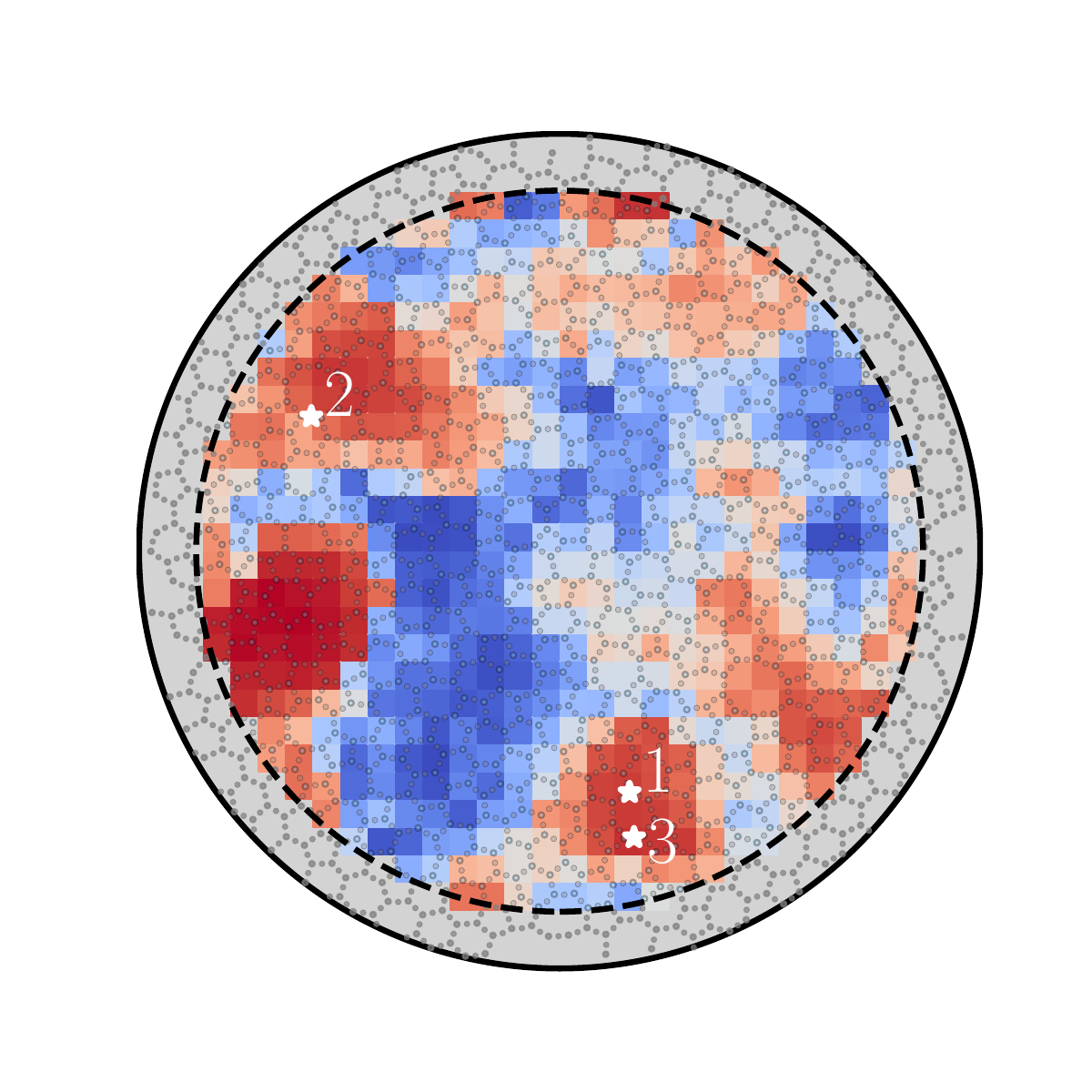}
        };
        \node at (0,4.6){\textbf{(a)}};
        \node at (0,-4.85){\textbf{(b)}};
        
    \end{tikzpicture}
    \caption{(a) Map of the predictor $\chi_{\!\hat{\bar{\sigma}}}$ of one $10^4$ atom sample; (b) Map of the predictor $\chi_{\!\hat{\bar{\sigma}}}$ to a real measured network glass topology. The actual occurring rearrangement events are indicated by white stars together with the sequence number at which they appear.}
    \label{fig:heatmap_results}
\end{figure}

Having exclusively focused on purely directional data which is mostly affected by the initial loading phase as shown in Figures \ref{fig:network_directions}b and \ref{fig:network_directions}c, we will now include information that considers the length of the network bonds. We define the random variable $\bar{\bm{n}}$ for the non-normalized bond vectors, collect the set of all its realizations $\bar{\bm{n}}^{(\alpha)}$ and define a random tensor $\bar{n}_{i_1}^{(\alpha)}\bar{n}_{i_2}^{(\alpha)}$. This way, the first moment, evaluated by $\bar{N}_{i_1i_2}:=\langle \bar{n}_{i_1}\bar{n}_{i_2}\rangle$, is the bond length extended equivalent to the normalized second order fabric tensor $N_{i_1i_2}$ above. However, we are interested in the second moment, that is, the variance of this random tensor, evaluated by:
\begin{align}
    \label{eq:variance_fabric_tensor}
    \bar{S}_{i_1i_2i_3i_4} :=&\langle (\bar{n}_{i_1}\bar{n}_{i_2} - \langle \bar{n}_{i_1}\bar{n}_{i_2} \rangle) (\bar{n}_{i_3}\bar{n}_{i_4} - \langle \bar{n}_{i_3}\bar{n}_{i_4} \rangle)\rangle \nonumber \\
    =& \langle \bar{n}_{i_1}\bar{n}_{i_2}\bar{n}_{i_3}\bar{n}_{i_4} \rangle - \bar{N}_{i_1i_2} \bar{N}_{i_3i_4} \; ,
\end{align}
which is a fourth-order tensor. This tensor contains the information of both anisotropy and variation in the bond length of the local network. Eigendecomposition leads to second-order eigentensors $\bm{\Phi}_i$ ($i=1,\dots,3$) and their corresponding eigenvalues $\lambda_i$. For further information we refer to the supplementary material. We define the predictor by $\chi_{\!\bar{S}} := \nicefrac{1}{\left(\lambda_1 |\bm{\Phi}_1 \cdot \bm{e}_1|\right)}$, which is the maximum variance of the bond lengths and its direction projected in the first principal direction of pure shear. The results of this predictor are shown by the purple plot in Figure \ref{fig:prediction_results}. One observes a huge improvement in performance from purely directional data, that is, unit orientation distribution functions, to directional and bond length data, that is, weighted orientation distribution functions. Based on these findings, one concludes that the bond length distribution plays a significant role in the mechanical deformation response of network glasses. Thus, we introduce one further predictor class, which is based on the concept of the bond length variance $\bar{\sigma} = \langle \left(\|\bar{\bm{n}}\| - \langle \|\bar{\bm{n}}\|\rangle\right)^2 \rangle$. The first predictor of this type is defined by $\chi_{\!\bar{\sigma}} := \nicefrac{1}{\bar{\sigma}}$ and presented by the black plot in Figure \ref{fig:prediction_results}. Surprisingly, the bond length variance shows high predictive performance even though it lacks any connection to the direction of the macroscopic deformation protocol. We finally modify the variance of the bond length variance by including directional information written as:
\begin{equation}
    \label{eq:bond_length_var_dir}
    \hat{\bar{\sigma}} := \langle \left(\|\bar{\bm{n}}\| - \langle \|\bar{\bm{n}}\|\rangle\right)^2 \bm{n}\bm{n}^{T} \rangle \cdot \bm{e}_1\bm{e}_1^{T} \; ,
\end{equation}
where we multiply every distance from the bond length expectation with the respective unit direction of the bond and project that vector into the principal direction of external pure shear deformation. Equivalently to before, the predictor is defined by $\chi_{\!\hat{\bar{\sigma}}} := \nicefrac{1}{\hat{\bar{\sigma}}}$. As shown in this figure, this directional bond length variance provides the most valuable scalar parameter of shear transformation zones. Figure \ref{fig:heatmap_results}a presents an example of a prediction map of $\chi_{\!\hat{\bar{\sigma}}}$ applied to one $10^4$ atom sample.
The blue regions indicate regions that are not expected to experience rearrangement events, while the red regions indicate shear transformation zones that are expected to be activated by external mechanical loading. We also localized the first eight events that occurred due to external deformation by white star-shaped markers together with their consecutive number of occurrences. Visual inspection reveals that all occurring events coincide with predicted shear transformation zones in the material. Even at larger strains where fracture is already quite progressed $\chi_{\!\hat{\bar{\sigma}}}$ shows high predictive performance.
To show the robustness and relevance of this predictor, we also included the topology from an image of another experimentally measured sample \cite{Huang2012}. We extracted a circular cutout of this sample of $63$ \r A and scanned the region using $\chi_{\!\hat{\bar{\sigma}}}$. To compare the predictor with actually activated regions, pure shear deformation was applied to the entire circular sample, but an outer circular area of $10$ \r A, which is equal to the cutoff radius of the potential, was excluded during the minimization at every athermal quasistatic deformation step. Due to the lack of periodic boundary conditions, we have only investigated the prediction of the first three events. At higher strains, the effects of the boundaries dominate the mechanical response. Also, for the measured sample, the actual event spots coincide with the predicted shear transformation zones. Visualization of the deformation of both the generated and the measured sample are presented in the supplementary material. There, it is shown that fracture indeed initiates in the predicted shear transformation zones.



This letter presents a new category of powerful predictors of shear transformation zones for network glasses. Compared to other indicators in the literature the predictors in this letter are purely geometrical; therefore, no expensive molecular simulations, eigenvalue analyses, or machine-learning enhanced strategies are required for a predictive assessment.
Having the structure of a covalent graph, network glasses respond to mechanical deformation like highly deformable truss structures, where the edges in the graph undergo considerable elastic rotations and stretching during the elastic range so that the geometrical picture shortly before an elementary event differs significantly from the initial state. However, our purely geometric predictors exclusively build on the general physical understanding of the mechanical response in the elastic range connected with actually occurring plastic instabilities. Furthermore, an important advantage of these indicators is that they can be directly measured in experiments.
The most crucial feature of network glass fracture is the local bond stretch in the network, quantified by the squared distance from the expected bond length.
Although this indicator can be further enhanced by information of bond anisotropy in relation to the direction of the loading protocol, leading to surprisingly high prediction accuracies, the most significant portion of prediction performance turns out to be invariant with respect to the direction of the deformation protocol.
We hope that the identification of local soft spots, described by molecular network systems of a few hundred degrees of freedom, by only one scalar parameter, which is largely rotationally invariant, paves the way to multiscale models of oxide glasses that take local atomic neighborhoods into account.


\begin{acknowledgments}
The authors acknowledge support from the German Research Foundation under the project number 523939420.
\end{acknowledgments}


\nocite{*}

\bibliography{lit}

\end{document}


\title{Geometrical fingerprints of shear transformation zones in network glasses \\ Supplementary material}
\date{\today}

\author{Franz Bamer}
\email{bamer@iam.rwth-aachen.de}
\affiliation{Institute of General Mechanics, RWTH Aachen University, Aachen, Germany}

\author{Zhao Wu}
\thanks{These two authors contributed equally}
\affiliation{Institute of General Mechanics, RWTH Aachen University, Aachen, Germany}
\author{Somar Shekh Alshabab}
\thanks{These two authors contributed equally}
\affiliation{Institute of General Mechanics, RWTH Aachen University, Aachen, Germany}

\maketitle

\section{Directional data}

\begin{figure}[hpt!]
\centering
\includegraphics{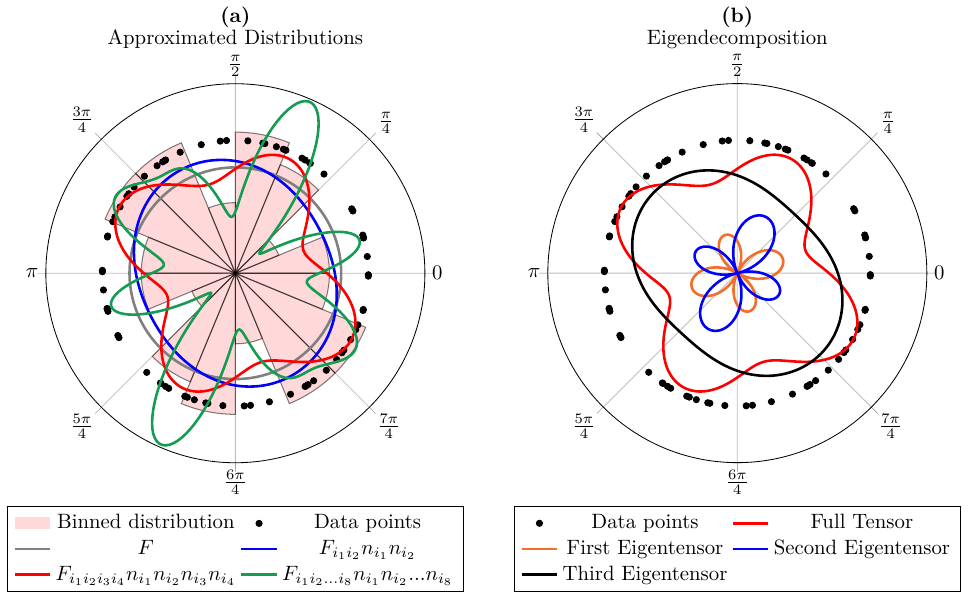}
\caption{Empirical and theoretical (fitted) distributions of normalized bond vector statistics. 
The empirical distribution is approximated using binning, while the theoretical distributions 
are approximated by the fabric tensor of the second kind. The gray line in the plot (a) represents the zeroth-order approximation, which corresponds to a circle. The blue line shows the second-order approximation, while the red and green lines correspond to the fourth- and eighth-order approximations, respectively. The plot in (b) shows the spectral decomposition of the fourth-order theoretical distribution function from plot (a). The red line presents the full tensor $F_{i_1i_2i_3i_4}$, whereas the green, blue, and black lines represent the contributions of the first, second, and third eigenmodes multiplied by the corresponding eigenvalues, respectively.}
\label{fig:dist_F}
\end{figure}

We define silica network structures as a graph \cite{Franzblau1991}.
In two dimensions, our model is an SiO$_{1.5}$ network glass \cite{Roy2018} consisting of corner-sharing SiO$_3$ triangles forming a network of rings of various shapes and sizes. Since our material is based on real images of flat silica structures \cite{Lichtenstein2012} we generate samples that are statistically equivalent to the benchmark images and are fully coordinated. The graph consists of nodes, that is, the centers of the SiO$_3$ units and edges, that is, connections of SiO$_3$ units with adjacent units. 
For an ensemble of particles, we extract the graph information and define a vector $\bm{n}$ which refers to the normalized direction between neighboring SiO$_3$ units. This vector is seen as a multivariate random variable. The empirical orientation distribution function is defined by:
\begin{equation}
    \phi(\bm{n}) = \frac{1}{N} \sum_{\alpha=1}^{N} \delta(\bm{n}-\bm{n}^{(\alpha)}) \; .
\end{equation}
Since $\phi(\bm{n})=\phi(-\bm{n})$ the orientation distribution function of the normalized bond vector is point symmetric. The moments of the orientation distribution function are written as:
\begin{equation}
    N_{i_1i_2...i_n} = \frac{1}{N} \sum_{\alpha=1}^{N} n^{(\alpha)}_{i_1} n^{(\alpha)}_{i_2}...n^{(\alpha)}_{i_n} = \langle n_{i_1} n_{i_2}...n_{i_n} \rangle \; .
\end{equation}
These moments, also referred to as fabric tensors of first kind, store the essential directional information. Due to the symmetry of the bond vectors, all fabric tensors of odd rank vanish.
To introduce a smooth, theoretical version of the orientation distribution function, we define the following tensor product:
\begin{equation}
    \phi(\bm{n}) \sim \frac{1}{4\pi}F_{i_1i_2....i_n} n_{i_1} n_{i_2} ...n_{i_n} \; ,
\end{equation}
where $F_{i_1i_2....i_n}$ is referred to as the fabric tensor of the second kind \cite{KENICHI1984}. Using the fabric tensors of first kind the fabric tensors of second kind are defined by:
\begin{equation}
        F_{i_1...i_n} = 2^n(N_{i_1...i_n} + a^n_{n-2}\delta_{(i_1i_2}N_{i_3...i_n)} + a^n_{n-4}\delta_{(i_1i_2}\delta_{i_3i_4}N_{i_5...i_n)} + ... + a^n_{0}\delta_{(i_1i_2}\delta_{i_3i_4}...\delta_{i_{n-1}i_n)}) \; .
\end{equation}
In this equation, the indices and the parameters $a^n_m$ and $c_n^m$ are defined as follows:
\begin{align}
    a^n_m =& \frac{1}{2^n}\sum^{n}_{\substack{k=m \\ k:\text{even}}}2^k c^k_{k-m} \; , \\
    c^n_m =& \frac{(-1)^{\frac{m}{2}}}{2^m}\frac{n}{n-\frac{m}{2}} \begin{pmatrix}
        n-\frac{m}{2} \\
        \frac{m}{2}
    \end{pmatrix} \; .
\end{align}
Clearly, the fabric tensors of second kind carry the same information as the fabric tensors of first kind; however, they provide an informative, smooth representation of the orientation function. Figure \ref{fig:dist_F} shows the histogram of the empirical orientation distribution function in gray.
Furthermore, this figure shows the second order fabric tensor of second kind in blue, the fourth-order fabric tensor of second kind in red and the eight-order fabric tensor of second kind in green.
The fabric tensors of second kind not only allow one to construct a smooth distribution function but also provide an intuitive physical interpretation of higher-order moments. A visual representative of the underlying geometry allows for a better understanding of the spatial organization and anisotropy in the material.

\section{Weighted directional data}

In the following, we incorporate bond stretching information into the tensorial quantities above. We consider non-normalized bond vectors, defined as $\bm{\bar{n}} = r\bm{n}$, where $r$ represents the bond length, that is, the absolute distance between neighboring triangle centers. This way, the bond vectors act as length-weighted multivariate random variables, encoding both directional and stretching information. Equivalently to above, the moments of the weighted orientation distribution function are written as:
\begin{equation}
    \bar{N}_{i_1i_2...i_n} = \langle \bar{n}_{i_1} \bar{n}_{i_2}...\bar{n}_{i_n} \rangle \; .
\end{equation}
This approach accounts for bond length and directional data, as well as their interactions. However, since the distributions are symmetric and all computed moments are centered around the zero vector, they do not directly capture fluctuations around the mean. To address this issue, we start from defining a length-weighted random bond tensor $B_{i_1i_2} := \bar{n}_{i_1} \bar{n}_{i_2}$ and an extended orientation distribution function $\bar{f}(B_{i_1i_2})$. The expectation of $B_{i_1i_2}$ is nonzero, and its moments are evaluated by:
\begin{equation}
    \bar{S}_{i_1i_2...i_{2n}} = \langle (\bar{n}_{i_1}\bar{n}_{i_2} - \langle \bar{n}_{i_1}\bar{n}_{i_2} \rangle)(\bar{n}_{i_3}\bar{n}_{i_4} - \langle \bar{n}_{i_3}\bar{n}_{i_4} \rangle)...(\bar{n}_{i_{2n-1}}\bar{n}_{i_{2n}} - \langle \bar{n}_{i_{2n-1}}\bar{n}_{i_{2n}} \rangle) \rangle \; .
\end{equation}
For example, the second-moment tensor of $\bar{f}(B_{i_1i_2})$ is written as:
\begin{equation}
    \bar{S}_{i_1i_2i_3i_4} = \langle (\bar{n}_{i_1}\bar{n}_{i_2} - \langle \bar{n}_{i_1}\bar{n}_{i_2} \rangle)(\bar{n}_{i_3}\bar{n}_{i_4} - \langle \bar{n}_{i_3}\bar{n}_{i_4} \rangle)\rangle = \bar{N}_{i_1i_2i_3i_4} - \bar{N}_{i_1i_2}\bar{N}_{i_3i_4} \; ,
\end{equation}
which quantifies the fluctuations in bond orientation and length, measuring how individual bonds deviate from the covariance $\langle \bar{n}_{i_1}\bar{n}_{i_2} \rangle$.
Furthermore, we introduce the distance of the bond length from the average bond length as a weighting function. In this case, the weighted fabric tensor takes the following form:
\begin{equation}
 \bar{\bar{N}}_{i_1i_2} = \frac{1}{N} \sum_{\alpha}^N \left( (r^{(\alpha)} - \langle r\rangle )^2 n^{(\alpha)}_{i_1}n^{(\alpha)}_{i_2} \right) \; ,
\end{equation}
which is equivalent to Equation (3) in the main text.

\subsection{Notes on the eigendecomposition of fabric tensors}

Due to the symmetry of the fourth-order fabric tensor $N_{i_1i_2i_3i_4}$, we use Kelvin's notation \cite{Sutcliffe1992} to rewrite the fourth objects as second-order tensors $\hat{N}_{i_1i_2}$. Equivalently, the second-order tensors are represented as a three-dimensional vectors. The basis vectors $\hat{\bm{e}}_i$ correspond to the combinations of the original basis vectors $\bm{e}_1 \otimes \bm{e}_1$-, $\bm{e}_2 \otimes \bm{e}_2$- and $\bm{e}_1 \otimes \bm{e}_2$. This allows one to express the fourth order fabric tensor in matrix form,
\begin{equation}
    \begin{aligned}
        \hat{N}_{i_1i_2} &= \begin{bmatrix}
            N_{1111} & N_{1122} &\sqrt{2}N_{1112} \\
            N_{2211} & N_{2222} & \sqrt{2}N_{2212} \\
            \sqrt{2} N_{1211} & \sqrt{2} N_{1222} & 2 N_{1212} \\
            \end{bmatrix} \; ,
    \end{aligned}
\end{equation}
and second-order tensors in vector form,
\begin{equation}
    \begin{aligned}
        \hat{\varphi}_{i} = \begin{bmatrix}
            \varphi_{11} \\
            \varphi_{22} \\
            2 \varphi_{12}
            \end{bmatrix} \; .
    \end{aligned}
\end{equation}
The linear mapping $\hat{N}_{i_1i_2}\hat{\varphi}_{i_2}$ is equivalent to $N_{i_1i_2i_3i_4}\varphi_{i_3i_4}$ when expressed in a two-dimensional Eucledian space. This enables the spectral decomposition of the tensor $N_{i_1i_2i_3i_4}$ in terms of its eigentensors as:
\begin{equation}
    N_{i_1i_2i_3i_4} = \sum^3_{\kappa=1} \lambda^{(\kappa)} \varphi_{i_1i_2}^{(\kappa)} \varphi_{i_3i_4}^{(\kappa)} \; ,
\end{equation}
where $\lambda^{(\kappa)}$ are the eigenvalues of the fourth-order tensor $N_{i_1i_2i_3i_4}$, and $\varphi_{i_1i_2}^{(\kappa)}$ are the corresponding second-order eigentensors.
For example, the fourth-order tensor $F_{i_1i_2i_3i_4}$ shown in Figure \ref{fig:dist_F}a, is decomposed in terms of its three eigentensors shown in \ref{fig:dist_F}b. The full tensor is presented in red, while the contributions from its eigenvalues are shown in orange, blue, and black, corresponding to the first, second, and third eigenvalues, respectively. The first eigenvalue exhibits a dipole shape, whereas the second and third eigenvalues display quadrupolar shapes. In fact, the second and third eigentensors are nearly deviatoric, which is inherited from the crystalline structure of silica network glass. In a crystalline honeycomb lattice, the first two eigentensors are purely deviatoric, with equal eigenvalues that are half the magnitude of the third eigenvalue.

\section{Geometrical predictors -- extended definitions and results}\label{sec:def_predictors}
For each spot, we compute the length-weighted fabric tensors up to the fourth order. Although the fabric tensors of second kind provide an intuitive and visual representation of the data, they do not offer additional insights that enhance soft spot prediction. Therefore, we focus our prediction analysis on the first-kind fabric tensors. 

\begin{figure}[hbt!]
    \centering
    \includegraphics[width=1.0\linewidth]{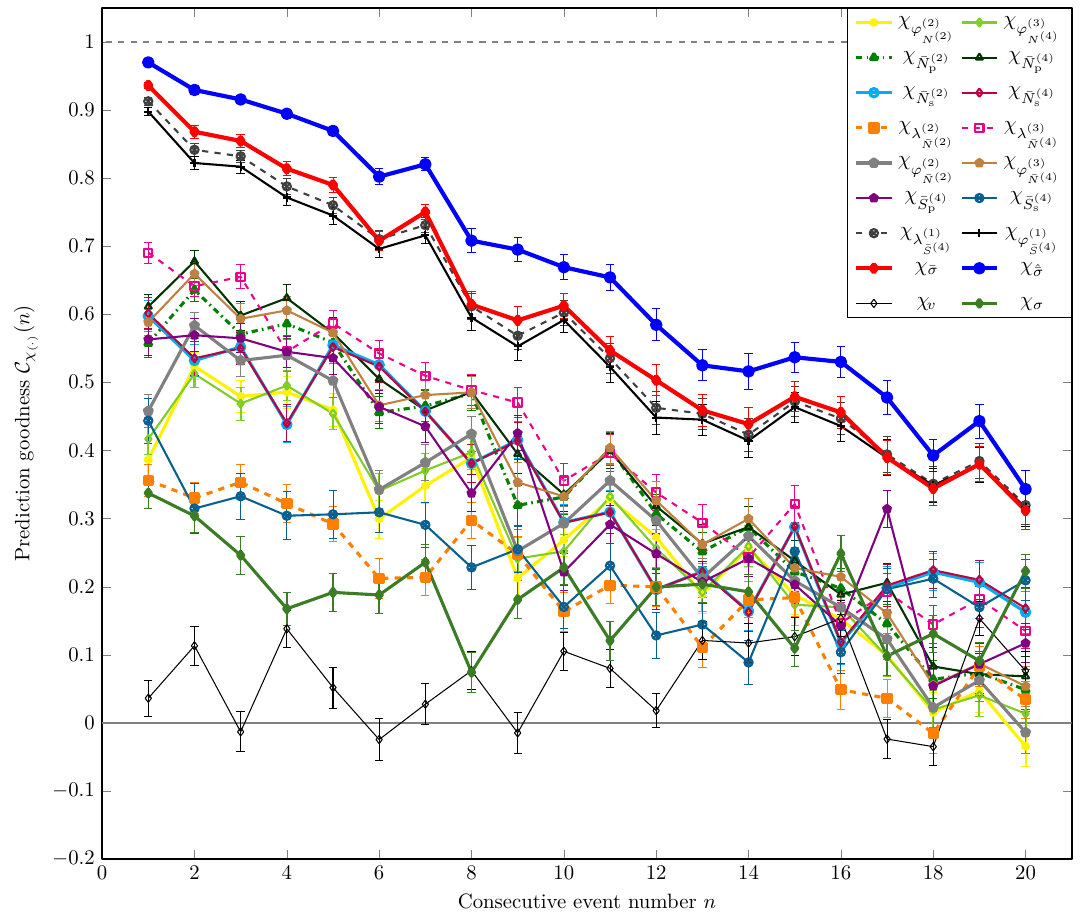}
    \caption{Prediction goodness of the computed indicators derived from the purely directional fabric tensors $N_{i_1i_2}$, $N_{i_1i_2i_3i_4}$, the bond-length-weighted fabric tensors $\bar{N}_{i_1i_2}$, $\bar{N}_{i_1i_2i_3i_4}$, and the extended fabric tensor $\bar{S}_{i_1i_2i_3i_4}$. Additionally, the indicators $\chi_{v}$, $\chi_{\bar{\sigma}}$, and $\chi_{\sigma}$, already presented in the main text, are included here. The error bars have been reduced by a factor of two for improved visibility. \textit{Note:} The indicator \raisebox{3pt}{\scalebox{0.8}{$\chi_{\varphi^{(1)}_{\bar{S}^{(4)}}}$}} is equivalent to $\chi_{\bar{S}}$ from the main text. However, the notation was modified here to facilitate comparison with other indicators derived from the tensor $\bar{S}_{i_1i_2i_3i_4}$.}
    \label{fig:goodness_sm}
\end{figure}

We primarily investigate five types of purely geometrical indicators: free volume, the local heterogeneity of the  ring statistics, the fabric tensors of the normalized bond vectors, the fabric tensors of length-weighted bond vectors, and measures related to the variance of bond length distributions. For each tensor, we extract scalar values which represent the higher dimensional objects and correlate with soft spots. Given a tensor $A_{i_1i_2...i_n}$, we define the following indicators:
\begin{itemize}
    \item Projection of the $n$-th order tensor $\bm{A}^{(n)}$ in the pulling direction of macroscopic pure shear deformation $\bm{e}_1$:
    \begin{equation}
        \chi_{A^{(n)}_\text{p}} = \frac{1}{\bm{A}^{(n)} \cdot (\bm{e}_1 \otimes \bm{e}_1 \otimes ... \otimes \bm{e}_1)}
    \end{equation}
    \item Projection of the $n$-th order tensor $\bm{A}^{(n)}$ in the pure shear direction of macroscopic deformation $\bm{F}_\text{s}$:
    \begin{equation}
        \chi_{A^{(n)}_\text{s}} = \frac{1}{\bm{A}^{(n)} \cdot (\bm{F}_\text{s} \otimes \bm{F}_\text{s} \otimes ... \otimes \bm{F}_\text{s})},
    \end{equation}
    where $\bm{F}_\text{s}$ is defined for pure shear deformation as:
    \begin{equation}
        \label{eq:deformation_gradient_pure_shear}
        \bm{F}_\text{s} = 
        \begin{pmatrix}
            1+\gamma & 0 \\
            0 & \frac{1}{1+\gamma}
        \end{pmatrix} \; ,
    \end{equation}
    where $\gamma$ refers to the shear strain parameter.
    \item The $m$-th eigenvalue of the $n$-th order tensor $\bm{A}^{(n)}$:
    \begin{equation}
        \chi_{\lambda^{(m)}_{\bm{A}}} = \frac{1}{|\lambda^{(m)}_{\bm{A}^{(n)}}|}
    \end{equation}
    \item The $m$-th eigenvector/tensor of the $n$-th order tensor scaled by its corresponding eigenvalue, projected in the pulling direction $\bm{e}_1$:
    \begin{equation}
        \chi_{\varphi^{(m)}_{A^{(n)}}} = \frac{1}{|\lambda^{(m)}_{\bm{A}^{(n)}}\bm{\varphi}^{(m)}_{\bm{A}^{(n)}} \cdot (\bm{e}_1 \otimes \bm{e}_1 \otimes ... \otimes \bm{e}_1)|}
    \end{equation}
\end{itemize}
We use the superscript $(\cdot)^{(n)}$ to denote the order of the tensor when applied to Latin letters and to refer to the eigenvalue order when applied to Greek letters. All prediction results are shown in Figure \ref{fig:goodness_sm}.

For the purely directional fabric tensor $N_{i_1i_2...i_n}$, we find that the eigenvalue-scaled eigenvector or tensor projected onto the pulling direction provides a relatively good prediction of soft spots. Projecting the tensor itself in the pulling direction yields similar results for the second-order tensor, but prediction goodness declines for higher orders. This can be attributed to stronger fluctuations in higher-order tensors due to data discreteness, leading to diminished goodness accuracy.

The inclusion of bond length in $\bar{N}_{i_1i_2...i_n}$ significantly enhances prediction goodness. Projection in the pulling direction provides better results than projection in the shear direction for the second-order tensor. Furthermore, the eigenvalue of the second-order tensor yields relatively poor predictions, whereas the eigenvector, when scaled by its eigenvalue and projected in the pulling direction, performs better. In all cases, prediction goodness improves with increasing tensor order, unlike the case for normalized fabric tensors, since the weighting function reduces the fluctuations.

For the tensor $\bar{S}_{i_1i_2...i_n}$, projection in the pulling direction outperforms the one in the shear direction, similar to other tensors. This suggests that the push direction is less relevant in network glasses. Finally, the eigenvalues and corresponding eigentensors provide significantly better prediction results, comparable to predictions based on bond length variance. Notably, the smallest eigenvalue appears to be highly correlated with the variance. 

\section{The influence of the scanning radius}

\begin{figure}[hbt!]
    \centering
    \includegraphics[width=1.0\linewidth]{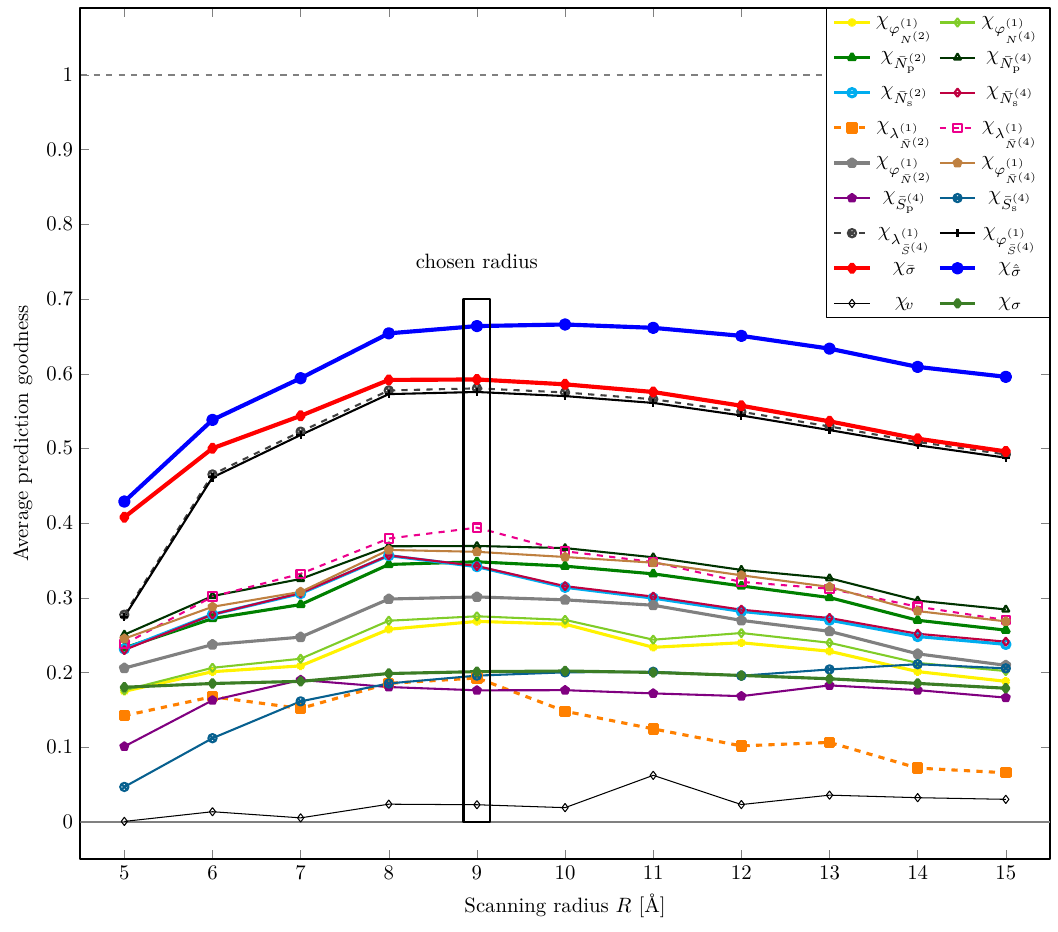}
    \caption{Average prediction goodness as a function of the scanning radius for all the indicators defined in \ref{sec:def_predictors} and shown in Figure \ref{fig:goodness_sm}.}
    \label{fig:goodness_rfree}
\end{figure}

For generating prediction maps, the samples are divided into a $50 \times 50$ grid. At each grid point, a circular region is defined to compute the local properties. The size of this region, determined by the scanning radius, is a critical parameter that must be carefully chosen since it provides essential information about the characteristic scale of the local material disorder, relevant for inelastic events. To evaluate its influence on the prediction performance, we compute the above-mentioned indicators changing the radius from $5.0$ to $15.0$ \AA. The average prediction goodness of the first $20$ events is computed for the variable scanning radius. The results are shown in Figure \ref{fig:goodness_rfree}. For most indicators, the prediction goodness increases significantly between $5.0$ \AA\ and $8.0$ \AA, it reaches its peak between $8.0$ \AA\ and $10.0$ \AA, and then gradually declines until a radius of $15.0$ \AA. Strictly speaking, there is no optimal collective scanning radius for all indicators. However, the value of $R = 9.0$ \AA appears to be the most suitable choice in general. Although larger radii incorporate more information, the averaging process suppresses essential local variations, ultimately resulting in a reduced prediction performance.

\section{Selected fracture simulations and their prediction}
\begin{figure}[hpt!]
    \centering
    \begin{tikzpicture}
        \node at (0,0){
            \includegraphics[width=0.8\linewidth]{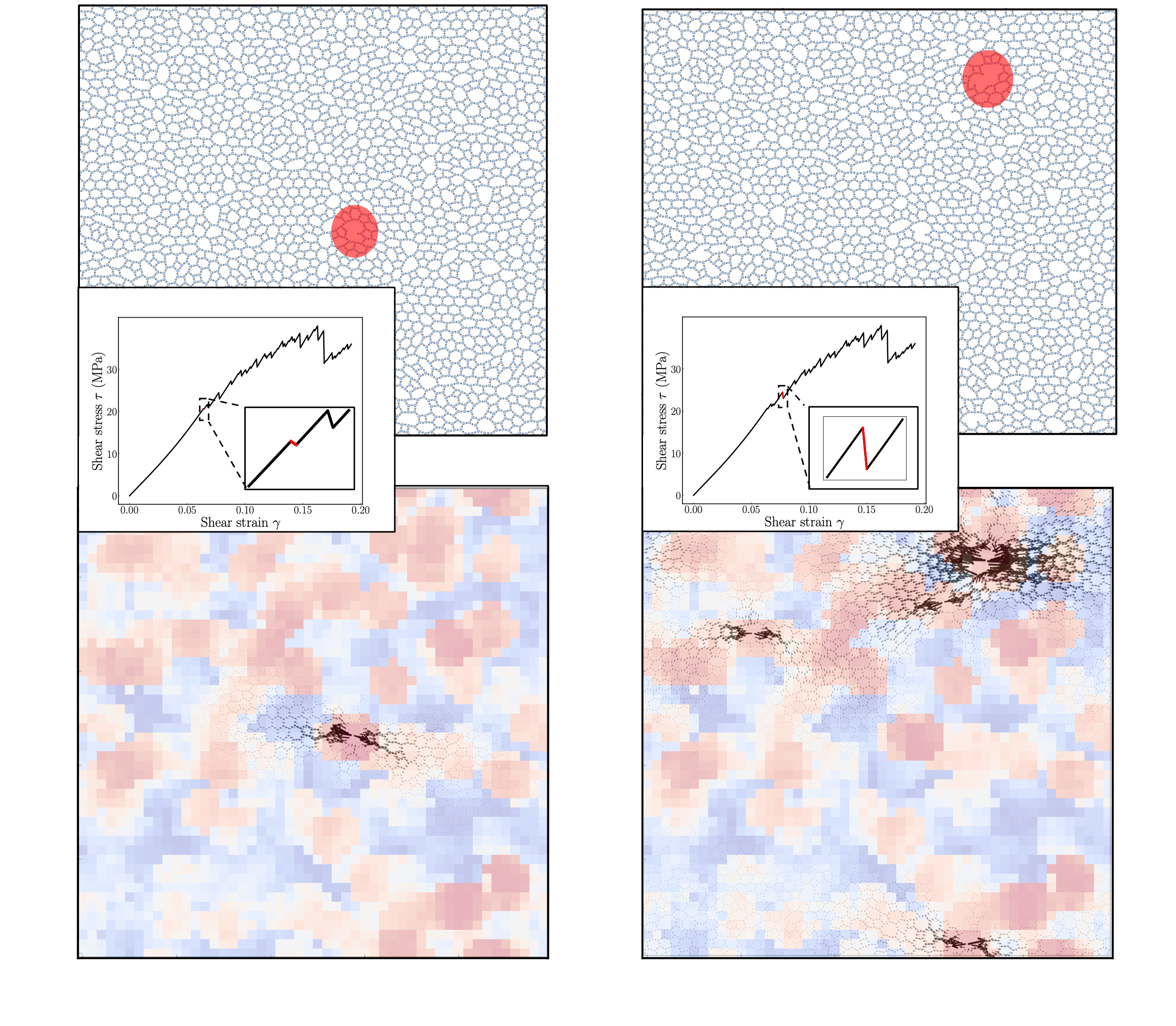}
        };
        \node at (-3.5,6.5){\textbf{(a)}};
        \node at (3.5,6.5){\textbf{(b)}};
    \end{tikzpicture}
    \caption{Macroscopic pure shear deformation of a two-dimensional network glass sample; two configurations of the first plastic event (a) and the fourth plastic event (b). The red circles highlight the local rearrangement caused by the plastic events and the corresponding stress drops are shown in the inlay of the stress-strain plots. The bottom row shows the corresponding nonaffine displacement field of each event, while the background presents the prediction map of $\chi_{\!\hat{\bar{\sigma}}}$.}
    \label{fig:deformation}
\end{figure}

The two-dimensional silica samples are subjected to pure shear deformation, which involves axial elongation in the first principal direction of shear and simultaneously compressing the second principal direction, as shown in Equation (\ref{eq:deformation_gradient_pure_shear}). Figure~\ref{fig:deformation} presents two snapshots captured at selected strain values, highlighting the local rearrangements during the deformation. The local rearrangements are indicated by red circles in the first row and the corresponding nonaffine displacement fields are shown in the bottom row. The bottom plots also show the  prediction heatmap $\chi_{\!\hat{\bar{\sigma}}}$. One observe the strong correlation between the actual regions of local rearrangements and the predicted areas likely to undergo rearrangements.

\begin{figure}[hpt!]
    \centering
    \begin{tikzpicture}
        \node at (0,0){
            \includegraphics[width=0.8\linewidth]{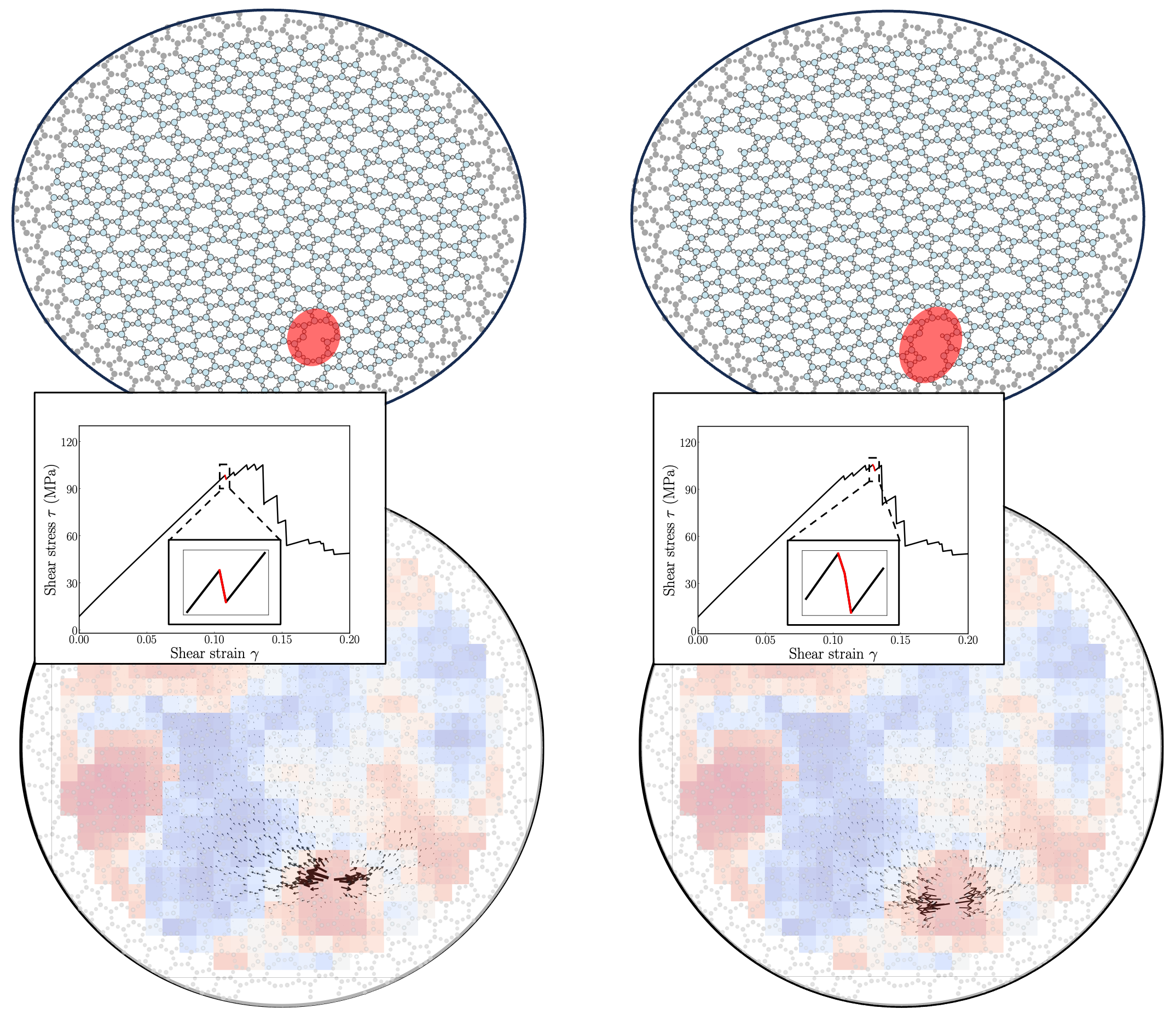}
        };
        \node at (-4.0,6.5){\textbf{(a)}};
        \node at (3.5,6.5){\textbf{(b)}};
    \end{tikzpicture}
    \caption{Macroscopic pure shear deformation of a real imaged 2D silica glass; two configurations of the first plastic event (a) and the fourth plastic event (b). The red circles highlight the local rearrangement caused by the plastic events and the corresponding stress drops are shown in the inlay of the stress-strain plots. The bottom row shows the corresponding nonaffine displacement field for the two subsequent events, while the background presents the prediction map $\chi_{\!\hat{\bar{\sigma}}}$.}
    \label{fig:deformation_real}
\end{figure}

Furthermore, we tested the prediction on an experimentally imaged real sample \cite{Huang2012} in Figure~\ref{fig:deformation_real}. As shown in this figure, the regions undergoing rearrangements align with the prediction prediction map $\chi_{\!\hat{\bar{\sigma}}}$.

\FloatBarrier
\bibliography{lit}